# RAMPING CONTROL OF HLS STORAGE RING

Jingyi Li, Gongfa Liu, Weimin Li, Chuan Li, Kaihong Li, Caozheng Diao, Zuping Liu
National Synchrotron Radiation Lab., P. O. Box 6022, Hefei, Anhui 230029, P. R. China

*Abstract*

HLS (Hefei Light Source) is a second generation synchrotron radiation source. After injected into the storage ring at energy of 200MeV, the electrons will be ramped to 800MeV. During ramping, the magnetic fields of the bend, quadruple and sextuple magnets are ramped up synchronously to keep the working point of the storage ring unchanged. This process is carefully controlled by HLS control system, which is built under EPICS. Several measures are taken to insure the synchronization and linearity of the ramping.

## 1 INTRODUCTION

HLS (Hefei Light Source) is a dedicated synchrotron light source. It consists of three parts, linac, transport line and storage ring [1]. After been accelerated to 200Mev by the linac, Electrons go through the transport line and then be injected into the storage ring. Because the working energy of HLS is 800Mev, the electrons need to be accelerated again to 800Mev, which called ramping. During ramping, the fields of all main magnets which includes bend, quadruple and sextupole magnets ramped up synchronously to keep the working point, i.e. bending radius, tune and chromaticity unchanged. This process is carefully controlled by HLS control system, which is built under EPICS [2]. The control of ramping includes three steps, ramping table calculation, ramping table downloading and ramping process control.

## 2 ABOUT RAMPING

For bend, quadruple and sextupole magnets, the integrated magnetic field can be described like following:

$$\begin{cases} B_{Bend} = \int_0^{L_B} B ds = \frac{(B\rho)}{\rho} L_B = \frac{E}{c\rho} L_B \\ B_Q = \int_0^{L_Q} \frac{\partial B_y}{\partial x} ds = (B\rho)|K|L_Q = \frac{E}{c} L_Q |K| \\ B_S = \int_0^{L_S} \frac{\partial^2 B_y}{\partial x^2} ds = (B\rho)|\lambda| = \frac{E}{c}|\lambda| \end{cases} \quad (1)$$

The bending radius ρ is determined by the energy and $B_{BEND}$. The tune is mainly determined by K, and chromaticity is mainly determined by λ. So the working point of the storage ring will not change if all of the bend, quadruple and sextupole magnets ramped up with the speed calculated with equation (1).

## 3 SYSTEM DESCRIPTION

Fig. 1 shows the control system of main magnets power supplies. Before ramping started, the ramping tables are calculated on the work station and then transferred to PS controller via the IOC. The ramping procedure is controlled by some records in the IOC. There is a interface to allow operators to control these records.

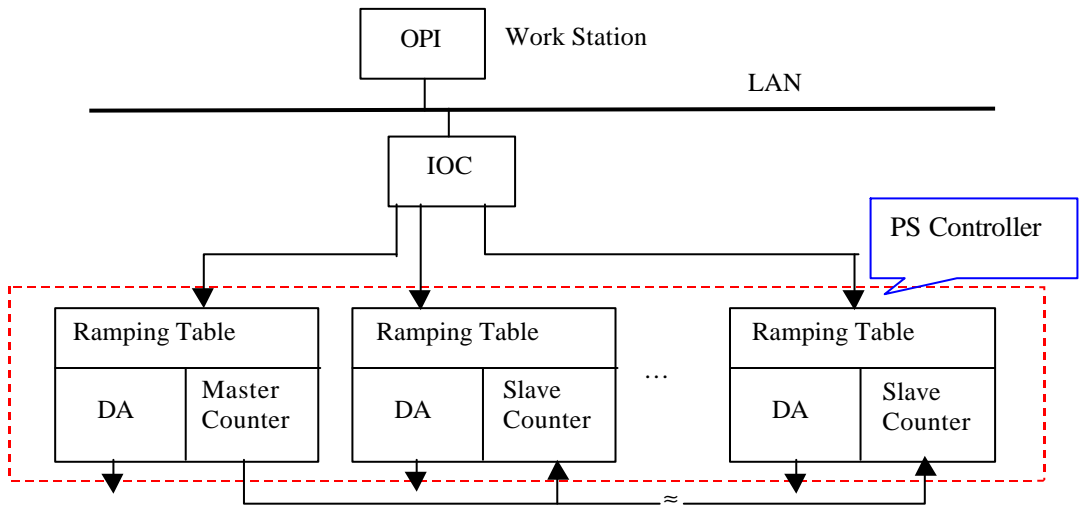

Fig. 1 HLS main magnet control system

## 4 RAMPING CALCULATION

Suppose the starting and end energy of ramping is $E_B$ and $E_E$. From equation (1), we get the corresponding $B_B$ and $B_E$. From the magnetic field measurements, we have the B-I relation. So we get the corresponding PS currents, $I_B$ and $I_E$. Suppose the time used during ramping is $\Delta t$, the speed of ramping is,

$$s = \frac{I_E - I_B}{\Delta t} \quad (2)$$

The PS controller control the PS setpoint by DAC, corresponding to the PS currents, the DAC value is $D_B$ and $D_E$, so the speed becomes,

$$s = \frac{\Delta D}{\Delta t} = \frac{D_E - D_B}{\Delta t} \quad (3)$$

Because the time is controlled by a counter, whose value together with the DAC value are all integers, the result of equation (3) must have an unwanted error. To minimize this error, let's suppose,

$$N_1 = (\text{int})\frac{\Delta t}{\Delta D}, N_2 = N_1 + 1 \quad (4)$$

$$\begin{cases} x + y = \Delta D \\ N_1 x + N_2 y = \Delta t \end{cases} \quad (5)$$

The solutions of equation (5) is,

$$\begin{cases} x = N_2 \Delta D - \Delta t \\ y = \Delta D - x = \Delta t - N_1 \Delta D \end{cases} \quad (6)$$

The maximum error is,

$$\boldsymbol{d}_{max} = \frac{xy}{\Delta t} \quad (7)$$

If $\delta_{max}$ bigger than the tolerance error $\delta$, we can divided x and y into (int) ($\delta_{max}/\delta$+1) parts, which can assure the maximum error is not bigger than the tolerance error $\delta$.

## 5  RAMPING CONTROL

Fig. 2 is the control panel of HLS ramping. When ramping is needed, it read the current setpoints of all of the main magnet power supplies. With a predefined configuration file, the program calculates ramping tables for these main magnet power supplies, and then sends them to corresponding PS controllers via the IOC. After the ramping table downloading is completed, the program sends a ramping preparation command to all of the PS controllers. As soon as get the ramping preparation command, the PS controller begins to prepare for ramping, which includes checking ramping table, spawning ramping task and waiting for step semaphore. After all of these work completed, the master counter begin to send signals to all of the slave counters. The slave counter counts these signals. When the count number reaches the corresponding number defined in the ramping table, the counter invoked an interrupt. There is an interrupt service routine for this interrupt, which will send a step semaphore to the ramping task. As soon as take the step semaphore, the ramping task increases the DAC value for 1 until it reaches the end of the ramping table.

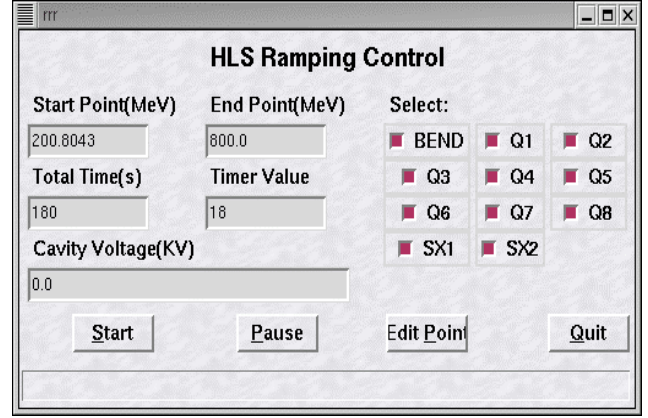

Fig. 2 The control panel of HLS ramping

## 6   CONCLUSION

This software has been running for about one year. According the experiences in the operation of HLS, the control of ramping is proven to be reliable.